\documentclass[review,sort&compress]{elsarticle}
\usepackage{amsmath}
\makeatletter
\renewcommand*\env@matrix[1][c]{\hskip -\arraycolsep
  \let\@ifnextchar\new@ifnextchar
  \array{*\c@MaxMatrixCols #1}}
\makeatother
\usepackage{amsfonts}
\usepackage[nolists,tablesfirst]{endfloat}
\usepackage{epstopdf}
\usepackage{setspace}
\usepackage{caption}
\usepackage{appendix}
\usepackage[colorlinks=true,citecolor=blue,linkcolor=blue]{hyperref}%

\begin{document}
\begin{frontmatter}
\journal{Acta Materialia}
\title{Thermodynamic Stability of Co-Al-W L1$_2$ $\gamma$'}

\author{James E. Saal\corref{cor1}}
\ead{j-saal@northwestern.edu}
\cortext[cor1]{Corresponding author}

\author{C. Wolverton\corref{dummy}}

\address{Department of Materials Science and Engineering,\\
Northwestern University, Evanston, IL 60208, USA}

\begin{abstract}
Co-based superalloys in the Co-Al-W system exhibit coherent L1$_2$ Co$_3$(Al,W) $\gamma$' precipitates in an fcc Co $\gamma$ matrix, analogous to Ni$_3$Al in Ni-based systems. Unlike Ni$_3$Al however, experimental observations of Co$_3$(Al,W) suggest that it is not a stable phase at 1173K. Here, we perform an extensive series of density functional theory (DFT) calculations of the $\gamma$' Co$_3$(Al,W) phase stability, including point defect energetics and finite-temperature contributions. We first confirm and extend previous DFT calculations of the metastability of L1$_2$ Co$_3$(Al$_{0.5}$W$_{0.5}$) $\gamma$' at 0K with respect to HCP Co, B2 CoAl, and D0$_{19}$ Co$_3$W using the special quasi-random structure (SQS) approach to describe the Al/W solid solution, employing several exchange/correlation functionals, structures with varying degrees of disorder, and newly developed larger SQS. We expand the validity of this conclusion by considering the formation of antisite and vacancy point defects, predicting defect formation energies similar in magnitude to Ni$_3$Al. However, in contrast to the Ni$_3$Al system, we find that substituting Co on Al sites is thermodynamically favorable at 0K, consistent with experimental observation of Co excess and Al deficiency in $\gamma$' with respect to the Co$_3$(Al$_{0.5}$W$_{0.5}$) composition. Lastly, we consider vibrational, electronic, and magnetic contributions to the free energy, finding that they promote the stability of $\gamma$', making the phase thermodynamically competitive with the convex hull at elevated temperature. Surprisingly, this is due to the relatively small finite-temperature contributions of one of the $\gamma$' decomposition products, B2 CoAl, effectively destabilizing the Co, CoAl, and Co$_3$W three phase mixture, thus stabilizing the $\gamma$' phase.
\end{abstract}

\end{frontmatter}

\newpage
\section{Introduction}
Ni-based superalloys have greatly advanced gas power and jet engine turbine efficiencies due to strong high-temperature mechanical performance\cite{reed2006superalloys}. This strength is commonly attributed to the $\gamma$/$\gamma$' FCC/L1$_2$ microstructure observed with Ni and Ni$_3$Al\cite{reed2006superalloys}. The discovery of similar $\gamma$/$\gamma$' high-temperature strengthening between FCC Co and L1$_2$ Co$_3$(Al,W)\cite{Sato2006} has made Co-based superalloys a promising avenue for further improvements in turbine blade performance as they have the potential to be stronger and melt at higher temperatures than Ni-based systems\cite{Pollock2010}. Interestingly, Co$_3$Al and Co$_3$W in the L1$_2$ structure are not stable phases, yet the observed L1$_2$ $\gamma$', consisting of a disordered solid solution of Al and W in the B site of the A$_3$B L1$_2$ structure, is readily observed\cite{Sato2006}. The reported compositions of $\gamma$' include Co-10Al-12W (at.\%)\cite{Sato2006} from field-emission electron probe microanalysis, nearly single-phase $\gamma$' cast from Co-10Al-12W and Co-10Al-13W\cite{Miura2007} and from Co-12Al-11W and Co-10Al-11W\cite{Inui2011}, and Co-10Al-12.5W from atom probe tomography\cite{Bocchini2012}. As such, the ``ideal'' $\gamma$' composition is often assumed to be equal amounts of Al and W mixing in the B sublattice, Co$_3$(Al$_{0.5}$W$_{0.5}$), although observed compositions tend to be slightly Co-rich and Al-poor relative to this ideal.

Understanding the thermodynamic stability of the $\gamma$/$\gamma$' two-phase system is vitally important for the long-term performance of a Co-based superalloy. The compositional width of the two-phase equilibrium in the ternary diagram (i.e. the degree to which the system can tolerate deviations in composition in either $\gamma$ or $\gamma$' before additional phases appear) is fairly narrow\cite{Sato2006}. As such, the optimization of high-temperature stability and mechanical properties with additional alloying elements has been aggressively investigated\cite{Pollock2010}. However, there is evidence that $\gamma$' in the Co-Al-W system itself may not be a thermodynamically stable phase. Diffusion couples between Co-27Al and Co-15W (at\%) initially form Co$_3$(Al,W) $\gamma$' along with FCC Co $\gamma$, B2 CoAl, and D0$_{19}$ Co$_3$W) after annealing at 1173K for 500 hours\cite{Kobayashi2009}. Annealing for 2000 hours, however, increases the amount of $\gamma$, CoAl, and Co$_3$W at the cost of $\gamma$', suggesting that $\gamma$' is metastable with respect to $\gamma$, CoAl, and Co$_3$W at 1173K. Similar results were observed in bulk Co-10Al-7.5W (at\%) samples, annealing at 1173K for 2000 hours\cite{Tsukamoto2010}.

Previous studies of Co$_3$(Al,W) $\gamma$' with first-principles density functional theory (DFT) calculations have explored the phase stability, predicting a metastable $\gamma$', with the disordered Co$_3$(Al$_{0.5}$W$_{0.5}$) L1$_2$ solid solution approximated by a 32 atom special quasi-random structure (SQS)\cite{Jiang2008,Mottura2012}. L1$_2$ Co$_3$(Al$_{0.5}$W$_{0.5}$) was found to be metastable by 70 meV/atom with respect to the three phase mixture of FCC Co, B2 CoAl, and D0$_{19}$ Co$_3$W\cite{Jiang2008,Mottura2012}. A negative formation energy of an ordered L1$_2$ supercell with respect to the elements was also predicted using WIEN2K\cite{Yao2006}. The effect of additional alloying elements has been investigated with DFT as well\cite{Chen2009a,Chen2010,Chen2010a,Mottura2012}. The DFT predicted single-crystal elastic constants of Co$_3$(Al,W) $\gamma$' are generally in good agreement with experiments\cite{Jiang2008,Wang2009a,Yao2006}.

Although the random solid solution of L1$_2$ Co$_3$(Al$_{0.5}$W$_{0.5}$) has been predicted to be metastable at 0K, there are other phenomena which can affect the stability of $\gamma$'. For instance, ordered structures will be more stable at 0K than the corresponding random solid solution at the same composition. However it is currently unknown how large this ordering energy is for L1$_2$ Co$_3$(Al$_{0.5}$W$_{0.5}$). Also, the composition of $\gamma$' is Co-rich and Al deficient with respect to the ``ideal'' Co$_3$(Al$_{0.5}$W$_{0.5}$)\cite{Sato2006,Miura2007,Inui2011,Bocchini2012}. Defects altering the composition must be present, such as Co vacancies or Al antisite defects, the energetics of which have not been calculated. Furthermore, finite-temperature phenomena such as vibrational effects must be considered to predict whether the 0K metastability extends to higher temperatures as $\gamma$' is observed in samples annealed at high temperature, on the order of 1200K.

As the stability of $\gamma$' is such a critical property, we revisit Co$_3$(Al$_{0.5}$W$_{0.5}$) $\gamma$' with DFT, exploring the above-mentioned phenomena which may affect the predicted metastability of the phase. We first confirm previous calculations of the metastability of L1$_2$ Co$_3$(Al$_{0.5}$W$_{0.5}$) $\gamma$' at 0K with respect to HCP Co, B2 CoAl, and D0$_{19}$ Co$_3$W with the SQS approach, employing several exchange/correlation functionals and structures with varying degrees of disorder, including a newly developed larger SQS. All calculations confirm 0K L1$_2$ Co$_3$(Al$_{0.5}$W$_{0.5}$ metastability. The effect of point defects on the stability is then studied, with the formation energies of antisite and vacancy-type defects calculated. We find similar defect formation energies as those for Ni$_3$Al, except that Co substitutions on Al sites are thermodynamically favored.  Lastly, finite-temperature contributions to the free energy of $\gamma$' are investigated, including vibrational, thermal electronic, and magnetic effects, all of which appear to promote $\gamma$' stability, primarily due to unique finite-temperature properties of B2 CoAl.

\section{Methodology}
Density functional theory calculations were performed with the Vienna Ab-initio Simulation Package (VASP)\cite{Kresse1996b,Kresse1996} using the included projector augmented wave potentials\cite{Kresse1999b}. The exchange and correlation functional was described with the generalized gradient approximation of Perdew, Burke, and Ernzerhof\cite{Perdew1996} (GGA-PBE). The 4s and 3d electrons of Co, the 3s and 3p electrons of Al, and the 5p, 6s, and 5d electrons of W were treated as valence, unless otherwise stated. Crystal structures were relaxed with respect to all degrees of freedom with an energy cutoff of 350 eV and gamma-centered k-point meshes of approximately 6000 k-points per reciprocal atom. All calculations are spin polarized.

The $\gamma$' L1$_2$ Co$_3$(Al,W) disordered solid solution is approximated by special quasi-random structures\cite{Zunger1990}, ordered supercells that reproduce the correlation functions of the disordered solution on average over all sites. SQSs have been constructed for a variety of common lattice types, including FCC\cite{Wolverton2001}, BCC\cite{Jiang2004}, HCP\cite{Shin2006}, and even the situation in the current paper, the B-sublattice of the A$_3$B L1$_2$\cite{Jiang2008}. Unless otherwise stated, we employ the previously published\cite{Jiang2011} L1$_2$ A$_3$(B,C) mixing SQSs, consisting of 8 mixing atoms (32 atoms total) for A$_3$(B$_{0.5}$C$_{0.5}$) and 16 mixing atoms (64 atoms total) for A$_3$(B$_{0.25}$C$_{0.75}$). In the current work, we identify SQSs by the number of mixing atoms, for instance SQS-8 contains 8 mixing atoms. We also generated an SQS-16 for A$_3$(B$_{0.5}$C$_{0.5}$) to better approximate the disordered solution and test the energetic convergence of the SQS with respect to supercell size, discussed in detail in the Appendix.

The 0K formation energy of a Co$_x$Al$_y$W$_z$ compound, $\Delta$E, is calculated by:
\begin{equation}\label{eqn:form}
\rm{\Delta E(Co_xAl_yW_z)=E(Co_xAl_yW_z)-\sum\limits_m\mu^m}
\end{equation}
where $\rm{E(Co_xAl_yW_z)}$ is the DFT total ground state energy of Co$_x$Al$_y$W$_z$ and $\rm{\mu^m}$ are the chemical potentials for Co, Al, and W, dictated by the choice of reference states. For the formation energy with respect to the elements, $\Delta$E$\rm{_f}$, the reference states are taken to be HCP Co, FCC Al, and BCC W. For the L1$_2$ mixing energy, $\Delta$E$\rm{_{mix}}$, the reference states are taken to be Co$_3$Al and Co$_3$W in the L1$_2$ structure. Of particular interest will be the stability of $\gamma$' L1$_2$ Co$_3$(Al$_{0.5}$W$_{0.5}$), $\Delta$E$\rm_{Stab}$, which we define by Equation \ref{eqn:form} with respect to the set of potentially stable binary and unary structures at this composition: HCP Co, D0$_{19}$ Co$_3$W, and B2 CoAl. If $\Delta$E$\rm_{Stab}$ is positive, then $\gamma$' is not stable. An equivalent relation can be constructed for finite-temperature contributions to the free energy of stability of $\gamma$' L1$_2$ Co$_3$(Al$_{0.5}$W$_{0.5}$), $\Delta$F$\rm{_{Stab}^{j}}$:
\begin{equation}\label{eqn:free-form}
\rm{\Delta F\rm{_{Stab}^{j}}[T]=F_j(Co_3(Al_{0.5}W_{0.5}))[T]-\sum\limits_m\mu^m_j[T]}
\end{equation}
where $\rm{F_j(Co_3(Al_{0.5}W_{0.5}))}$ is the contribution from physical effect j to the free energy of L1$_2$ Co$_3$(Al$_{0.5}$W$_{0.5}$) and $\rm{\mu^m_j}$ is the contribution to the chemical potential due to physical effect j. If $\Delta$F$\rm{_{Stab}^{j}}$ is negative then j has a stabilizing effect on $\gamma$'. In this work, the vibrational and thermal electronic contributions to the free energy are calculated. The vibrational contribution to the free energy is predicted by frozen phonon calculations with the supercell approach in the harmonic approximation\cite{Wolverton2004,Wolverton2007a}. Supercell sizes are given in Table \ref{tab:phonon}. The thermal electronic free energy is predicted with the Alloy Theoretic Automated Toolkit (ATAT)\cite{VandeWalle2009}.

Antisite and vacancy point defects are considered in this work, designated by X$\rm_Y$, where X is the species on the Y site of the pristine L1$_2$. $\square$ indicates a vacancy. The defect formation energies, $\Delta$E$\rm{_{Defect}(X_Y)}$, of antisite defects and vacancies in the 32 atom SQS-8 structure, Co$_{24}$Al$_4$W$_4$, are calculated by:
\begin{equation}\label{eqn:defect_co-al}
\rm{\Delta E_{Defect}=E(Defect)-E(Pristine)-\sum\limits_m\mu^m}
\end{equation}
where E(Defect) is the energy of the supercell containing the point defect, E(Pristine) is the energy of the perfect Co$_{24}$Al$_4$W$_4$ SQS-8 structure. The chemical potentials are taken to be HCP Co, B2 CoAl, and D0$_{19}$ Co$_3$W.

\section{Results and Discussion}

\subsection{SQS 0K Stability}
\begin{figure}[tbp]
\centering %
\includegraphics[width=4in]{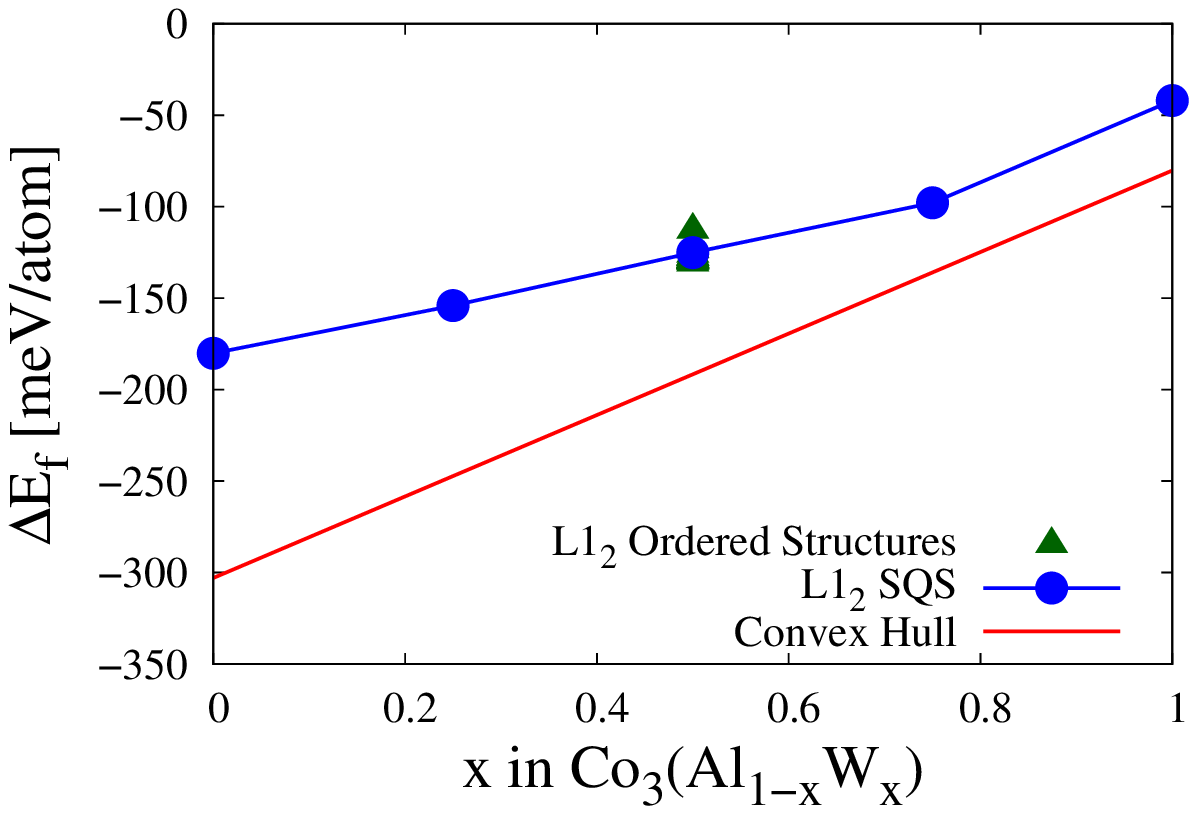}\\
\includegraphics[width=4in]{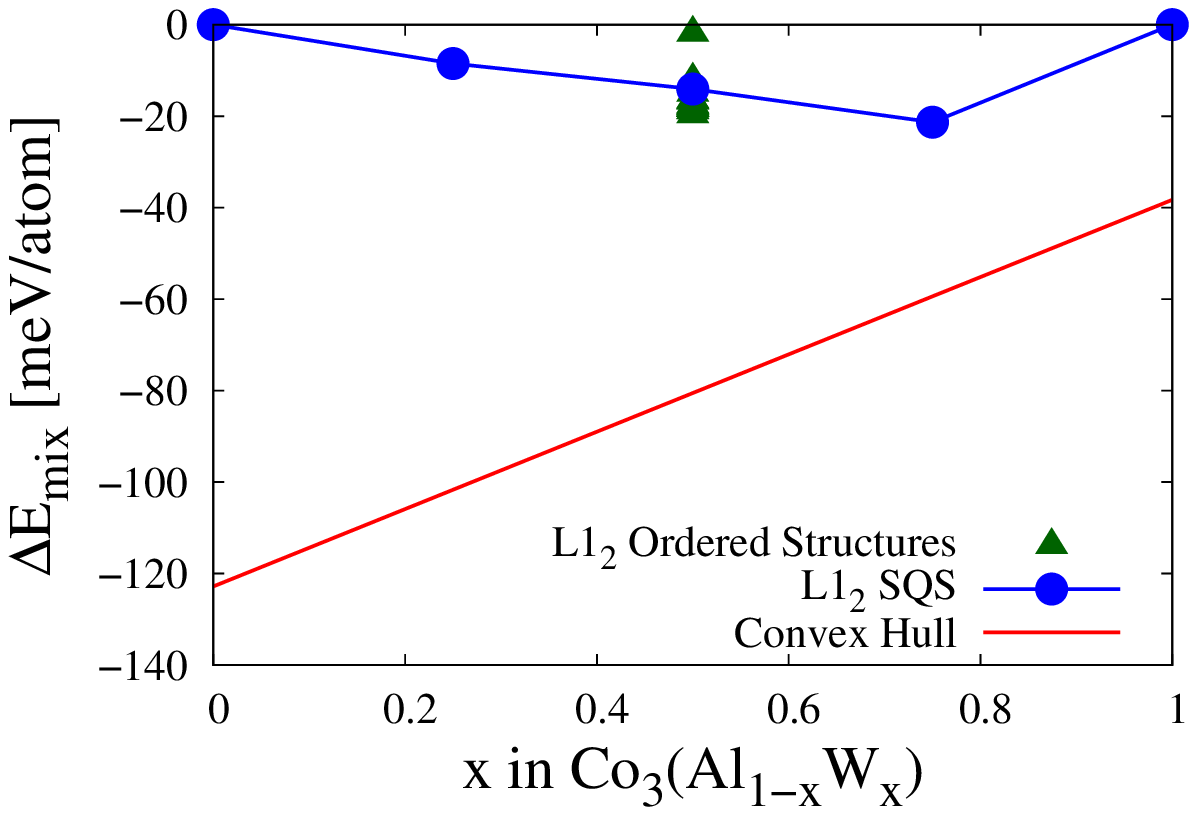}\\
\caption{DFT formation energies with respect to the elements (top) and mixing energies with respect to the Co$_3$Al and Co$_3$W L1$_2$ structures (bottom) for the L1$_2$ Co$_3$(Al$\rm_{1-x}$W$\rm_x$) SQSs, the L1$_2$ ordered structures given in Table \ref{tab:hf-struct}, and the Co$_3$Al-Co$_3$W convex hull, consisting of HCP Co+B2 CoAl+D0$_{19}$ Co$_3$W.}
\label{fig:chull}
\end{figure}

The formation energy with respect to the elements, $\Delta$E$\rm{_f}$, and the L1$_2$ mixing energy, $\Delta$E$\rm{_{mix}}$, of the L1$_2$ Co$_3$(Al$\rm_{1-x}$W$\rm_x$) disordered solid solution from the SQS approach are shown in Figure \ref{fig:chull}. Also shown are the energies of a series of Al/W ordered L1$_2$ supercells (discussed later) and the predicted thermodynamically stable set of phases at these compositions (the so-called convex hull), consisting of the HCP Co+B2 CoAl+D0$_{19}$ Co$_3$W three-phase mixture. Although the Co$_3$(Al$\rm_{1-x}$W$\rm_x$) SQSs are thermodynamically stable with respect to the elements, the solid solution is not stable with respect to the other phases in the Co-Al-W system at 0K. The formation energy of Co$_3$(Al$_{0.5}$W$_{0.5}$) SQS-8, -125 meV/atom, lies 66 meV/atom ($\Delta$E$\rm{_{Stab}}$) above the convex hull. In other words, Co$_3$(Al$_{0.5}$W$_{0.5}$) L1$_2$ is predicted to transform to HCP Co+B2 CoAl+D0$_{19}$ Co$_3$W at 0K.  The SQS results are nearly identical to those of previous work\cite{Jiang2011,Mottura2012}. To put the magnitude of this $\Delta$E$\rm{_{Stab}}$ in context, the configurational free energy at 1200K due to ideal Al/W mixing is only -18 meV/atom, given by $-k_BT\frac{1}{4}(\frac{1}{2}\ln\frac{1}{2}+\frac{1}{2}\ln\frac{1}{2})$. Configurational entropy alone will not make Co$_3$(Al$_{0.5}$W$_{0.5}$) more stable than HCP Co+B2 CoAl+D0$_{19}$ Co$_3$W at 1200K.  1200K will be used as the given temperature for all free energy contributions as this is approximately the temperature where $\gamma$' was first observed\cite{Sato2006} and the annealing temperature for the experiments where $\gamma$' was observed to be metastable\cite{Kobayashi2009,Tsukamoto2010}.

DFT's ability to predict the stability of $\gamma$' can be estimated by comparing to experimental formation energies of binary compounds. However, the only experimentally measured $\Delta$E$\rm{_f}$ in the Co-Al and Co-W binary systems is for B2 CoAl, determined calorimetrically to be -561$\pm$30\cite{Hultgren1973,Oelsen1937} and -694\cite{Blitz1937} meV/atom. Since the current DFT predicted value (-606 mev/atom) lies between the experiments and the range between the experiments is so large (130 meV/atom or 20\% of the DFT value), an assessment of DFT's accuracy in predicting stability in the Co-Al-W system is difficult from these experimental values.

To explore the possibility that the metastability of the L1$_2$ Co$_3$(Al$_{0.5}$W$_{0.5}$) is an artifact of some error in the choice of GGA-PBE for the exchange and correlation functional, $\Delta$E$\rm{_{Stab}}$ for SQS-8 was calculated with different exchange and correlation functionals: GGA-PW91\cite{Perdew1992} and the local density approximation (LDA) of Ceperley and Alder\cite{Ceperley1980} as parameterized by Perdew and Zunger\cite{Perdew1981}. We also tested a W potential with fewer valence electrons (the 5p electrons are placed in the core) with GGA-PBE. The resulting stabilities are given in Table \ref{tab:hf-pot}. The GGA calculations give similar values, varying by only 3 meV/atom. Switching from GGA to LDA, on the other hand, has a much larger effect, reducing $\Delta$E$\rm{_{Stab}}$ by 19 meV/atom. However, regardless of DFT settings, Co$_3$(Al$_{0.5}$W$_{0.5}$) SQS-8 is still predicted to be metastable, suggesting that \textit{the prediction of metastability is independent of the exchange and correlation functional}.

\begin{table}[tbp]
  \centering
  \caption{Energetic stability of L1$_2$ Co$_3$(Al$_{0.5}$W$_{0.5}$) SQS-8 with respect to the convex hull, in meV/atom, for various exchange and correlation functional and W potentials.}
    \begin{tabular}{lc}
    \hline\hline
 & $\Delta$E$\rm{_{Stab}}$\\
    \hline
    GGA-PBE             & 66 \\
    GGA-PW91            & 63 \\
    LDA                 & 47 \\
    GGA-PBE (W core 5p) & 65 \\
    \end{tabular}%
  \label{tab:hf-pot}%
\end{table}

A key assumption in the current calculations is that the disordered L1$_2$ Co$_3$(Al$_{0.5}$W$_{0.5}$) solid solution is adequately represented by the SQS-8 structure. To test this assumption, $\Delta$E$\rm{_{Stab}}$ for structures of varying degrees of disorder were predicted. We have included calculations of the smallest ordered structure at the Co$_3$(Al$_{0.5}$W$_{0.5}$) composition, a 2$\times$1$\times$1 L1$_2$ supercell containing alternating planes of Al and W (henceforth referred to as the 211 structure), and twenty other ordered supercells of varying sizes and Al/W arrangements. We also constructed an SQS twice as large as SQS-8, SQS-16, with correlation functions much closer to those of a random solid solution (see Appendix). The resulting $\Delta$E$\rm{_{Stab}}$ values, shown in Table \ref{tab:hf-struct}, range by only 18 meV/atom, suggesting that the effect on the total energy due the chemical arrangement of Al and W on the B-sublattice of the A$_3$B L1$_2$ is somewhat screened by the Co on the A-sublattice. None of the structures are predicted to be stable relative to the convex hull, confirming the prediction of $\gamma$' metastability at 0K is not an artifact of the degree of order in the system. The smallest structure is also the most energetically favorable, 5 meV/atom more stable than SQS-8. This result is consistent with the expectation that ordered compounds are energetically favored over disordered solid solutions at 0K. The temperature above which the Al/W sublattice will energetically prefer the disordered state over the ordered state, from thermal energy alone ($k_BT$), is 58K (i.e. above this temperature the 5 meV/atom difference between the ordered and disordered Co$_3$(Al$_{0.5}$W$_{0.5}$) will be overcome by the thermal energy), suggesting that the Al/W sublattice is indeed disordered at the high temperatures $\gamma$' is observed.

\begin{table}[tbp]
  \centering
  \caption{Energetic stability, in meV/atom, of L1$_2$ Co$_3$(Al$_{0.5}$W$_{0.5}$) structures with varying supercell size and disorder, with the GGA-PBE potential.}
    \begin{tabular}{lcc}
    \hline\hline
    Structure & \# Al/W & $\Delta$E$\rm{_{Stab}}$  \\
    \hline
    211 & 2     & 61 \\
    20 Other Ordered & 4-8   & 62-79 \\
    SQS-8            & 8     & 66 \\
    SQS-16           & 16    & 67 \\
    \end{tabular}%
  \label{tab:hf-struct}%
\end{table}

\subsection{Defects}
As previously stated, relative to the current SQS calculation at Co-12.5Al-12.5W, the observed compositions of $\gamma$' are typically Co-rich and Al-poor\cite{Sato2006,Miura2007,Inui2011,Bocchini2012}, suggesting defects are present to deviate from the ``ideal'' Co$_3$(Al$_{0.5}$W$_{0.5}$ composition discussed earlier. We investigate deviation from this ideal composition by way of antisite defects and vacancies in the SQS-8. These defects have the potential to further energetically stabilize the Co$_3$(Al$_{0.5}$W$_{0.5}$) SQS while also introducing additional configurational entropy.

Table \ref{tab:defects} summarizes the defect formation energies, energetic stabilities, and the configurational free energy (including both defect and Al/W mixing) at 1200K. A range of energies are given in Table \ref{tab:defects} since the SQS-8 contains Al, W, and Co sites of varying local chemistry. All possible defect sites were calculated in the current work (e.g. all 24 Co atoms in the SQS-8 were individually replaced with Al for the Co$\rm{_{Al}}$ defect). Antisite defects include replacing Al and W with Co, Co$\rm{_{Al}}$ and Co$\rm{_W}$, respectively, and replacing Co with Al and W, Al$\rm{_{Co}}$ and W$\rm{_{Co}}$, respectively.  Notably, Co$\rm{_{Al}}$ has a negative defect formation energy, whereas all the other defects are positive. Vacancies of Al, Co, and W ($\square\rm{_{Al}}$, $\square\rm{_{Co}}$, and $\square\rm{_{W}}$, respectively) were also calculated. All three vacancy types are predicted to be thermodynamically unfavorable, with defect formation energies larger than 1000 meV/atom. The formation energies of both the antisite and vacancy point defects (except for Co$\rm{_{Al}}$) are of similar magnitude to their equivalent in Ni$_3$Al\cite{Fu1997}. Also note that the configurational free energy from these defects contributes to the free energy. As an example, for both Al/W mixing and a single $\square\rm{_{Co}}$ defect in SQS-8 the configurational free energy is calculated by $-k_BT\frac{3}{4}(\frac{1}{24}\ln\frac{1}{24}+\frac{23}{24}\ln\frac{23}{24})-k_BT\frac{1}{4}(\frac{1}{2}\ln\frac{1}{2}+\frac{1}{2}\ln\frac{1}{2})$. With point defects, the configurational contribution to the free energy at 1200K increases from -18 meV/atom for only ideal Al/W mixing to, at most, -31 meV/atom. Single site defect calculations were performed on the SQS-16 structure as well to test whether the SQS-8 supercell size is sufficiently large to isolate the point defects and minimize artificial defect-defect interactions due to the periodic boundary conditions employed in DFT.  The SQS-16 defect formation energies (shown in parentheses in Table \ref{tab:defects}) are similar to those of SQS-8, indicating the current SQS-8 defect formation energies are converged with respect to supercell size.

\begin{table}[tbp]
  \centering
  \caption{Energetics of antisite defects and vacancies in Co$_3$(Al$_{0.5}$W$_{0.5}$) SQS-8 (and SQS-16 in parentheses), including defect formation energy, energetic stability with respect to the convex hull, and the configurational free energy (including both defect and Al/W mixing) at 1200K. NCo$\rm_{Al}$ refers to N additional substitutions of Al with Co, and their corresponding $\Delta$E$\rm{_{Defect}}$ is for the cost of placing the additional defect.}
    \begin{tabular}{lccc}
    \hline\hline
          & $\Delta$E$\rm{_{Defect}}$ & $\Delta$E$\rm{_{Stab}}$ & F$\rm{^{Config}_{1200K}}$  \\
          & [meV/defect]              & [meV/atom]              & [meV/atom]\\
    \hline
    Co$\rm{_{Al}}$  & -93 - 25 (-115)   & 64 - 67  & -25 \\
   2Co$\rm{_{Al}}$ & -91 - 63   & 61 - 66  & -27 \\
   3Co$\rm{_{Al}}$ &  101 - 102 & 64 - 64  & -25 \\
   4Co$\rm{_{Al}}$ & 162        & 69       & -18 \\
    Co$\rm{_W}$     & 275 - 423 (285) & 75 - 80  & -25 \\
    Al$\rm{_{Co}}$  & 263 - 825  (409) & 75 - 92  & -31 \\
    W$\rm{_{Co}}$   & 300 - 1464 (857) & 76 - 112 & -31 \\
    \\
    $\square\rm{_{Al}}$ & 1996 - 2001 (2034) &133 - 133& -25 \\
    $\square\rm{_{W}}$  & 2255 - 2438 (2312) &141 - 147& -25 \\
    $\square\rm{_{Co}}$ & 1476 - 1932 (1637) &116 - 131& -31 \\
    \end{tabular}%
  \label{tab:defects}%
\end{table}

The negative defect formation energy for Co$\rm{_{Al}}$ indicates that replacing Al with Co stabilizes the SQS-8 energetically with respect to the convex hull at 0K, perhaps approaching a more stable Co-rich ordered compound. This suggests $\gamma$' may prefer Co-rich compositions from the ideal Co$_3$(Al$_{0.5}$W$_{0.5}$), in agreement with the observed $\gamma$' compositions\cite{Sato2006,Miura2007,Inui2011,Bocchini2012}. Furthermore, the observed W composition is very similar to the ideal 12.5 at.\%, consistent with the current predictions since Co$\rm{_W}$ is not as competitive as Co$\rm{_{Al}}$. To explore the limits of Co$\rm{_{Al}}$ favorability, additional Al atoms were substituted with Co (indicated as 2Co$\rm{_{Al}}$, 3Co$\rm{_{Al}}$, and 4Co$\rm{_{Al}}$ for two, three, and four Al replacements, respectively). The results are given in Table \ref{tab:defects}, where the defect formation energies are taken with respect to the previous Co$\rm{_{Al}}$ defect. Interestingly, the second Co$\rm{_{Al}}$ replacement is also favorable, by about the same energy as the first. The third and fourth, however, are no longer favorable. Although Co$\rm{_{Al}}$ defects are favorable, their effect on the stability of $\gamma$' is small, reducing the metastability by less than 10 meV/atom for 2Co$\rm{_{Al}}$. In the A$_3$B L1$_2$ structure, a B-site has six nearest neighbor B-sites. Al-sites that correspond to favorable Co$\rm{_{Al}}$ substitutions in the SQS-8 happen to be those that are rich in Al in these six nearest neighbor B-sites, indicating Co-Al L1$_2$ nearest neighbor B-site bonds are favored.

\subsection{Finite-Temperature Effects}
Beyond configurational disorder, other finite-temperature contributions can alter the stability of $\gamma$' at the elevated temperatures in which it is experimentally observed, including vibrational, electronic, and magnetic effects. Of these phenomena, phonons, the characteristic vibrations of a crystalline lattice, can have a large impact on the stability of metal alloys\cite{Ozolins1998,Wolverton2001a,Walle2002,Mao2011a,Mao2011b}. If the vibrational properties of $\gamma$' are sufficiently different from the competing phases, then there is the possibility the vibrational free energy of stability, $\Delta$F$\rm{_{Stab}^{Vib}}$, will be negative, having a stabilizing effect on $\gamma$'. $\Delta$F$\rm{_{Stab}^{Vib}}$ for Co$_3$(Al$_{0.5}$W$_{0.5}$) is predicted from DFT by calculating, with the frozen phonon supercell approach, the vibrational free energies of HCP CO, B2 CoAl, D0$_{19}$ Co$_3$W, and L1$_2$ Co$_3$(Al$_{0.5}$W$_{0.5}$) for the 21l and SQS-8 structures. The vibrational free energies of these structures at 1200K are given in Table \ref{tab:phonon}. The 16 atom Co vibrational free energy is less than 1\% different from the 54 atom value, indicating 54 atoms is sufficiently converged for the current work. Interestingly, the difference between the SQS-8 and 211 ordered vibrational free energies is also very small, suggesting that the vibrational properties of the chemical ordering of the B sublattice in A$_3$B L1$_2$ are screened by the A sublattice. $\Delta$F$\rm{_{Stab}^{Vib}}$ at 1200K is -27 meV/atom, contributing a stabilizing effect to Co$_3$(Al$_{0.5}$W$_{0.5}$). The negative $\Delta$F$\rm{_{Stab}^{Vib}}$ is due to the vibrational free energy of CoAl, which is smaller than those of the other structures, indicating stiffer bonds than the other structures. A calculation of the CoAl vibrational entropy with a larger supercell results in a similar value (also given in Table \ref{tab:phonon}), suggesting the CoAl vibrational free energy is converged with respect to supercell size.

\begin{table}[tbp]
  \centering
  \caption{Vibrational free energies, in meV/atom, at 1200K predicted by DFT with the supercell phonon approach for Co-Al-W structures.}
    \begin{tabular}{lcc}
    \hline\hline
          &\# Atoms& F$\rm{^{Vib}_{1200K}}$  \\
    \hline
HCP Co  & 16  & -451\\
        & 54  & -449 \\
B2 CoAl & 54  & -361 \\
        & 128 & -365\\
D0$_{19}$ Co$_3$W &64 & -443\\
\\
L1$_2$ Co$_3$(Al$_{0.5}$W$_{0.5}$) &    & \\
211                 & 54 & -451 \\
SQS-8                              & 32 & -453 \\
    \end{tabular}%
  \label{tab:phonon}%
\end{table}

\begin{figure}[tbp]
\centering %
\includegraphics[width=3in]{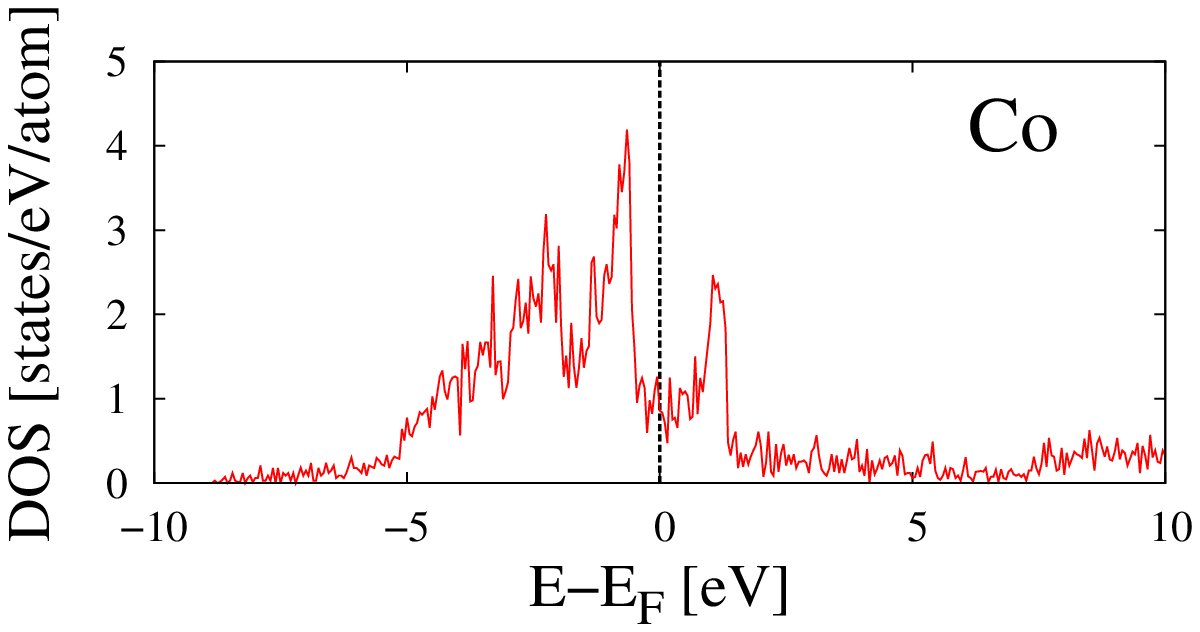}\\
\includegraphics[width=3in]{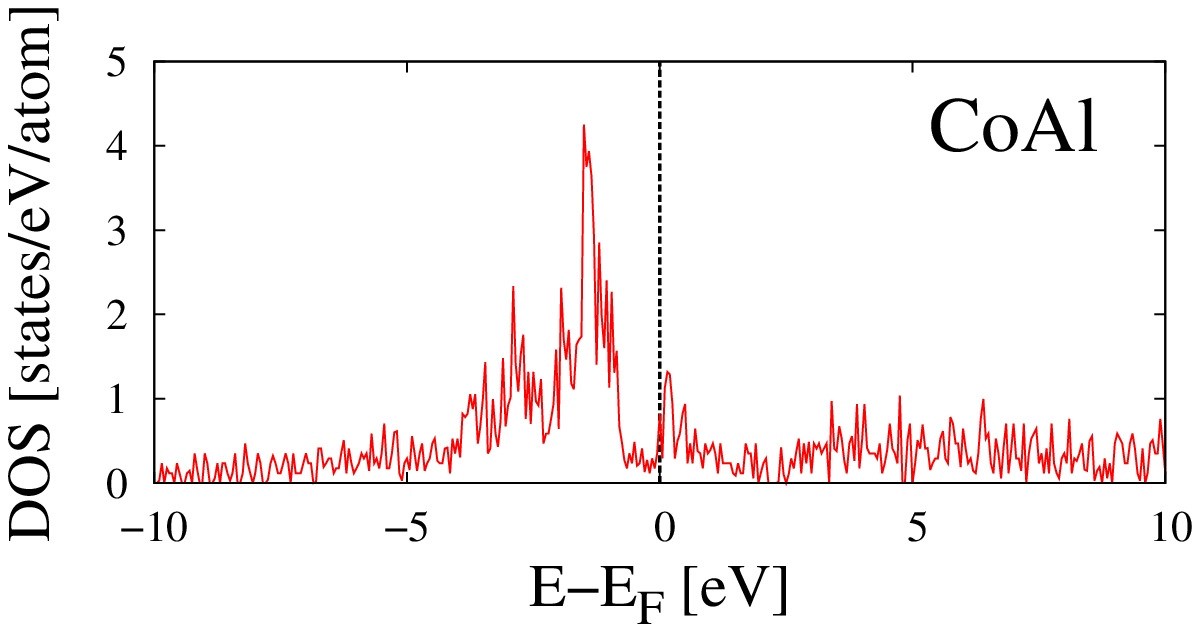}\\
\includegraphics[width=3in]{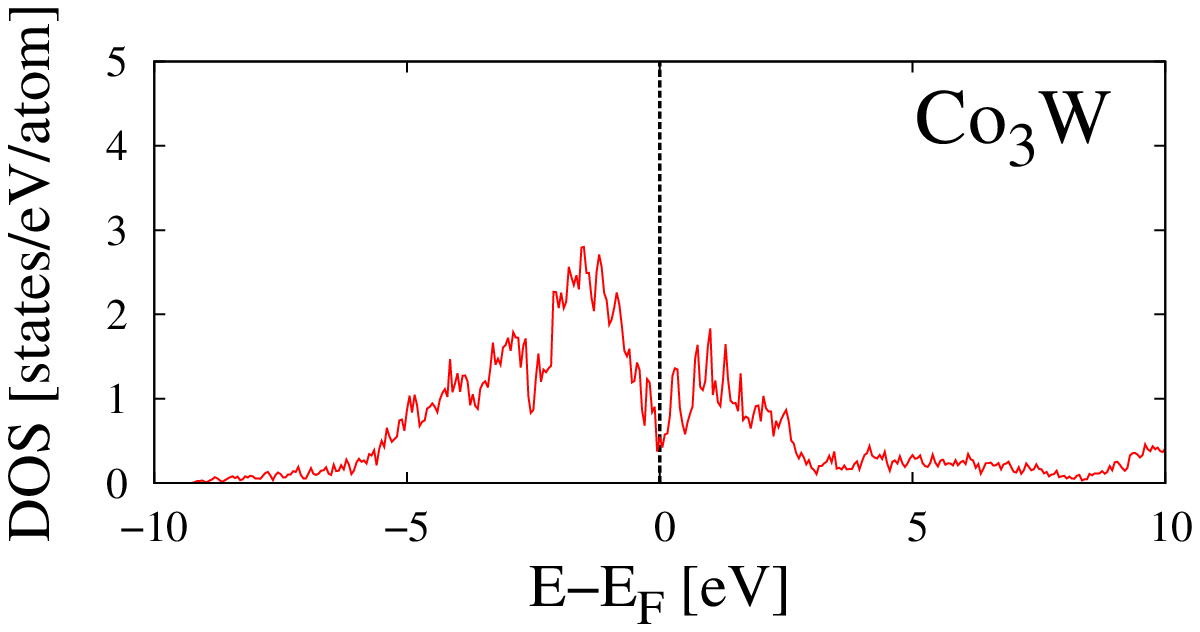}\\
\includegraphics[width=3in]{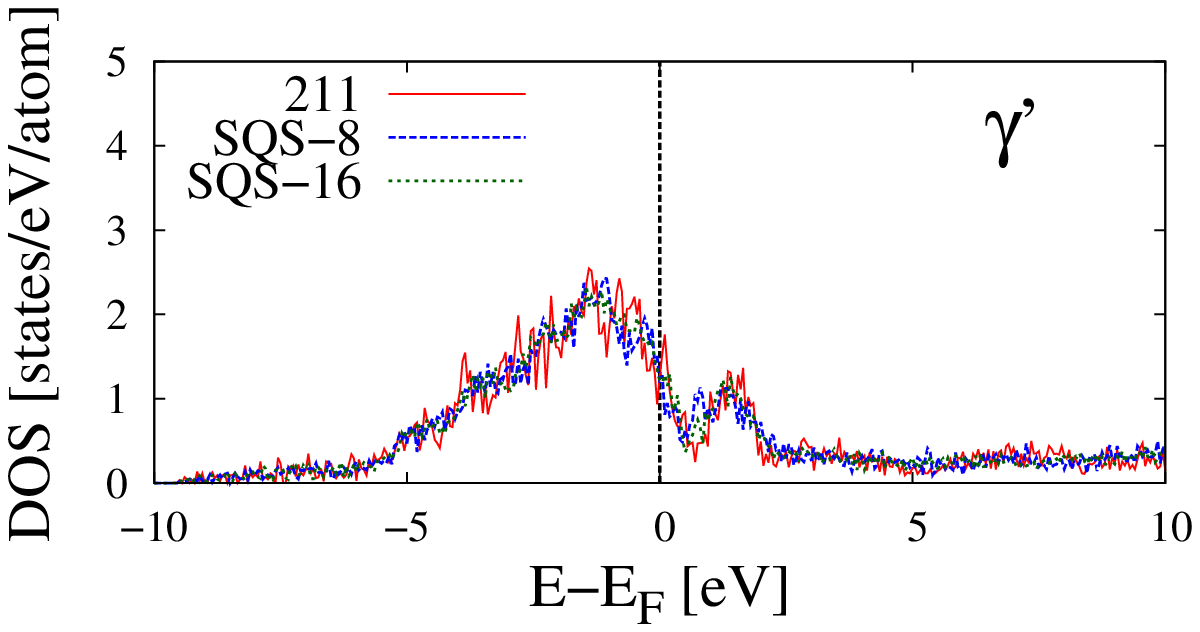}\\
\includegraphics[width=3in]{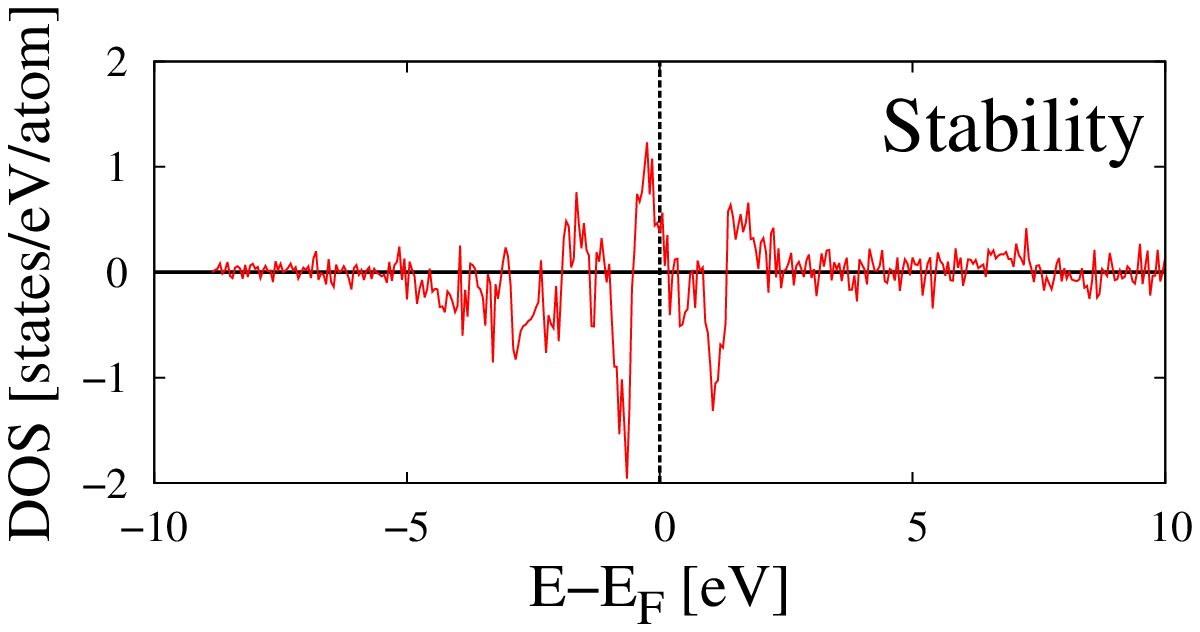}\\
\caption{Total electronic densities of states for HCP Co, B2 CoAl, D0$_{19}$ Co$_3$W, and L1$_2$ Co$_3$(Al$_{0.5}$W$_{0.5}$) with the 211, SQS-8, and SQS-16 structures. The difference between the L1$_2$ Co$_3$(Al$_{0.5}$W$_{0.5}$) SQS-8 DOS and the other phases is shown as a DOS of stability, calculated by taking the weighted difference of the DOSs as in $\Delta$E$\rm{_{Stab}}$. Energies are normalized to the Fermi level (E$\rm{_F}$).}
\label{fig:edos}
\end{figure}

Another possible contribution to the free energy is the thermal excitation of the electrons from their 0K ground state below the Fermi level to excited states. This contribution must be considered for metals since the electronic density of states (DOS) will be non-zero at the Fermi level\cite{Wolverton1995}. The electronic DOS, predicted from DFT, for HCP Co, B2 CoAl, D0$_{19}$ Co$_3$W, and L1$_2$ Co$_3$(Al$_{0.5}$W$_{0.5}$) with the 211, SQS-8, and SQS-16 structures are shown in Figure \ref{fig:edos}. Electronic excitations will affect $\gamma$' stability if the $\gamma$' DOS near the Fermi level is different from those of the competing phases. This difference is calculated by taking the weighted difference of the DOSs as in $\Delta$E$\rm{_{Stab}}$ (effectively a DOS of stability), using SQS-8 for Co$_3$(Al$_{0.5}$W$_{0.5}$). Also shown in Figure \ref{fig:edos}, the DOS of stability is non-zero near the Fermi level, particularly just below it.  This peak in the DOS of stability is due to B2 CoAl, which exhibits a deep well in the DOS below the Fermi level, in agreement with previous calculations\cite{Rhee2000}. Calculation of the electronic free energies at 1200K, F$\rm{^{El}_{1200K}}$, are given in Table \ref{tab:felec} and confirm the imbalance between $\gamma$' and the other structures. The DOSs for the L1$_2$ Co$_3$(Al$_{0.5}$W$_{0.5}$) structures show good agreement with previous calculations\cite{Jiang2008,Yao2006}, and the difference in the electronic free energies and DOSs between the three structures is minor, suggesting that the arrangement of Al and W atoms has little effect on the electronic free energy. The electronic free energy of stability, $\Delta$F$_{Stab}^{El}$, calculated from the SQS-8 electronic free energy, is -10 meV/atom. In other words, the electronic free energy also has a stabilizing effect on Co$_3$(Al$_{0.5}$W$_{0.5}$).

\begin{table}[tbp]
  \centering
  \caption{Electronic free energy at 1200K predicted by DFT, in meV/atom.}
    \begin{tabular}{lcc}
    \hline\hline
          & F$\rm{^{El}_{1200K}}$  \\
    \hline
HCP Co  &-16\\
B2 CoAl & -8\\
D0$_{19}$ Co$_3$W & -12\\
\\
Co$_3$(Al$_{0.5}$W$_{0.5}$) &\\
211 & -23\\
SQS-8 &-22 \\
SQS-16 &-23 \\
    \end{tabular}%
  \label{tab:felec}%
\end{table}

Accurate prediction of magnetic thermodynamics at elevated temperatures with DFT requires sophisticated models\cite{Kormann2008,Shang2010c,Wang2008b,Shang2010b,Pajda2001,Kuzmin2006,Kuzmin2005} beyond the scope of the current work. However, some general comments can be made from the current 0K ground state calculations. As with vibrational effects, magnetism can affect the stability of $\gamma$' if the any of the phases it competes with have sufficiently different magnetic properties. Co, Co$_3$W, and $\gamma$' SQS-8 are predicted to have ferromagnetic order on the Co sites, with average magnetic moments of 1.61, 0.31, 0.74 $\mu\rm{_B}$, respectively. CoAl, however, is diamagnetic in DFT, with no local Co moment. This is an experimentally observed phenomenon\cite{Rhee2000}, attributed to the Al valence electrons pairing with the Co valence electrons responsible for magnetism. Therefore, the magnetic free energy of stability of $\gamma$' is most likely negative, as CoAl has no magnetic entropy to contribute towards stabilizing the HCP Co+B2 CoAl+D0$_{19}$ Co$_3$W three-phase mixture. An upper bound on the sensitivity of $\Delta$E$\rm{_{Stab}}$ due to magnetism can be estimated by calculating the stability from DFT total energies without spin moments considered, essentially the nonmagnetic $\gamma$' 0K stability. We predict this value to be 12 meV/atom, 56 meV/atom lower than the magnetic value, indicating that magnetism has the potential to greatly affect to the stability of $\gamma$'.

Taken altogether, the calculated contributions to the free energy of stability for $\gamma$' L1$_2$ Co$_3$(Al$_{0.5}$W$_{0.5}$) are predicted to promote the stability of the phase. At 1200K, the free energy contribution from the ideal configurational entropy of Al/W mixing is -18 meV/atom, -27 meV/atom from vibrations, and -10 meV/atom from thermal electronic excitations. These effects reduce the metastability of the SQS-8 structure from 66 meV/atom above the convex hull to only 11 meV/atom. Additional stabilization from defects and magnetism will likely reduce it further, possibly making $\gamma$' thermodynamically competitive with the other phases in the Co-Al-W system.

\section{Summary}
The DFT prediction of $\gamma$' L1$_2$ Co$_3$(Al$_{0.5}$W$_{0.5}$) metastability has been explored in detail. Various structures with different degrees of disorder and several DFT exchange/correlation functionals agree that L1$_2$ Co$_3$(Al$_{0.5}$W$_{0.5}$) lies above the convex hull of HCP Co, B2 CoAl, and D0$_{19}$ Co$_3$W at 0K, with the PBE SQS-8 prediction as metastable by 66 meV/atom. Substitutional point defects have small positive defect formation energies while vacancies are much more unfavorable. However, substituting Al sites with Co is predicted to be thermodynamically favorable at 0K, in agreement with experimental measurements of excess Co and Al deficiency in $\gamma$' relative to the ideal Co$_3$(Al$_{0.5}$W$_{0.5}$) composition. Finite-temperature contributions to the free energy have been predicted as well. Lattice vibrations and electronic excitations appear to have a stabilizing contribution to the free energy of L1$_2$ Co$_3$(Al$_{0.5}$W$_{0.5}$) due to small free energy contributions from B2 CoAl. Magnetism may also promote $\gamma$' stability as CoAl is nonmagnetic. Summing all the predicted contributions to the free energy of $\gamma$' reduces the metastability from 66 to 11 meV/atom. $\gamma$' may become less metastable, possibly even stable, when considering further finite-temperature contributions, such as magnetism.

\section*{Acknowledgements}
This research was sponsored by the US Department of Energy, Office of Basic Energy Sciences (Dr. John Vetrano, monitor) through grant DE-FG02-98ER45721. Calculations were performed on the Northwestern University high performance computing system, Quest. Many thanks to David Seidman, David Dunand, and Peter Bocchini at Northwestern University and Carelyn Campbell, Ursula Kattner, and Eric Lass at NIST for fruitful discussions.

\newpage
\section*{Appendix: Simple Cubic SQS-16}
\label{sec:appendix}

The disordered solid solution on the B-site in the A$_3$B L1$_2$ of $\gamma$' is described with special quasi-random structure approach\cite{Zunger1990}. The B-site of L1$_2$ is equivalent to a simple cubic lattice. An SQS for this sublattice at A$_3$(B$_{0.5}$C$_{0.5}$) composition containing 8 mixing atoms (SQS-8) has previously been developed and employed to test the stability of Co$_3$(Al,W) $\gamma$'\cite{Jiang2008}. To assess the ability of this SQS-8 to predict the enthalpy of mixing for $\gamma$', we generate a simple cubic SQS with twice the mixing sites, SQS-16, which will serve as a better approximation for the disordered solid solution.

We begin by enumerating all possible SQS structures of up to 16 mixing atoms (360,642 structures) using a lattice enumeration algorithm\cite{Hart2008}. The pair, triplet, and quadruplet correlation functions, a measure of the degree of chemical disorder in a fixed lattice, for all 360,642 structures were calculated with the corrdump function of ATAT\cite{VandeWalle2009}. Of all possible structures, a single structure, detailed in Table \ref{tbl:sqspos}, has pair correlations identical to that of the random solution out to the 14th nearest neighbor shell. The next best structure approximates the random solutions only as far as the 6th. In comparison, the published SQS-8 deviates from the random solid solution pair correlation function by the 4th nearest neighbor shell. On this basis, the structure in Table \ref{tbl:sqspos} was chosen as the SQS-16 used in the DFT calculations in the current work. The pair, triplet, and quadruplet correlation functions for the new SQS-16, the published SQS-8, and the ideal disordered solid solution are given in Table \ref{tbl:sqscf}. The new SQS-16 performs better than the published SQS-8 at approximating the pair and quadruplet correlation functions of the disordered solid solution, whereas similar triplet correlation performance is present in the two SQSs.

\begin{table}[p]
\centering%
\caption{Crystal structure of the simple cubic SQS-16 developed and employed in the current work.}
\label{tbl:sqspos} %
\begin{tabular}{ccccccc}
\hline \hline
\multicolumn{7}{c}{Lattice Vectors
$\begin{pmatrix}[r]
 1 & 0 & 0 \\
-1 & 2 & 1 \\
-3 &-2 & 3 \\
\end{pmatrix}$
}  \\
\multicolumn{3}{c}{Element 1} && \multicolumn{3}{c}{Element 2}\\
x     & y     & z      & &  x     & y     & z   \\
\cline{1-3} \cline{5-7}
0.000	&	0.000	&	0.000	&&	0.500	&	0.625	&	0.625	\\
0.500	&	0.125	&	0.125	&&	0.000	&	0.750	&	0.750	\\
0.000	&	0.250	&	0.250	&&	0.500	&	0.875	&	0.875	\\
0.500	&	0.375	&	0.375	&&	0.500	&	0.500	&	0.000	\\
0.000	&	0.500	&	0.500	&&	0.000	&	0.625	&	0.125	\\
0.500	&	0.750	&	0.250	&&	0.000	&	0.875	&	0.375	\\
0.000	&	0.125	&	0.625	&&	0.500	&	0.000	&	0.500	\\
0.000	&	0.375	&	0.875	&&	0.500	&	0.250	&	0.750	\\
\end{tabular}
\end{table}

\begin{table}[htbp]
\centering %
\caption{Simple cubic correlation functions, $\overline{\Pi}_{k,m}[n]$, of pairs $(k=2)$, triplets $(k=3)$, and quadruplets $(k=4)$ at the $m$th nearest neighbor shell, each with $n$ equivalent figures. Correlation functions are provided for the random solid solution, the SQS-16, and the previously published SQS-8\cite{Jiang2008}.}
\label{tbl:sqscf}
\begin{tabular}{cccccccccccc}
\hline \hline%
  & Rnd. &16&8 & &Rnd. &16 &8& & Rnd. &16&8 \\
\hline
$\overline{\Pi}_{2,1}$[6] & 0 & 0&0&$\overline{\Pi}_{3,1}$[12] & 0 & 0&0&$\overline{\Pi}_{4,1}$[3] & 0 & -1/3&-1/3         \\
$\overline{\Pi}_{2,2}$[4] & 0 & 0&0&$\overline{\Pi}_{3,1}$[8] & 0 & 0&0&$\overline{\Pi}_{4,1}$[8] & 0 & 0&0         \\
$\overline{\Pi}_{2,3}$[3] & 0 & 0&0&$\overline{\Pi}_{3,2}$[24] & 0 & 0&0&$\overline{\Pi}_{4,1}$[2] & 0 & 0&1        \\
$\overline{\Pi}_{2,4}$[12] & 0 & 0&-1/3&$\overline{\Pi}_{3,3}$[3] & 0 & 0&0&$\overline{\Pi}_{4,2}$[24] & 0 & 0&0  \\
$\overline{\Pi}_{2,5}$[12] & 0 & 0&0&$\overline{\Pi}_{3,3}$[12] & 0 & 0&0&$\overline{\Pi}_{4,2}$[24] & 0 & -1/12&0         \\
$\overline{\Pi}_{2,6}$[6] & 0 & 0&0&$\overline{\Pi}_{3,3}$[12] & 0 & 0&0&$\overline{\Pi}_{4,2}$[6] & 0 & 1/6&-1/3 \\
$\overline{\Pi}_{2,7}$[3] & 0 & 0&-1/3&$\overline{\Pi}_{3,4}$[24] & 0 & 0&0&$\overline{\Pi}_{4,3}$[12] & 0 & 0&0 \\
$\overline{\Pi}_{2,7}$[12] & 0 & 0&0&$\overline{\Pi}_{3,4}$[24] & 0 & 0&0&$\overline{\Pi}_{4,3}$[12] & 0 & -1/6&0 \\
$\overline{\Pi}_{2,8}$[12] & 0 & 0&0&$\overline{\Pi}_{3,4}$[48] & 0 & 0&0&$\overline{\Pi}_{4,3}$[3] & 0 & 1/3&-1/3 \\
$\overline{\Pi}_{2,9}$[12] & 0 & 0&0&$\overline{\Pi}_{3,4}$[24] & 0 & 0&0&$\overline{\Pi}_{4,3}$[12] & 0 & 1/6&0 \\
$\overline{\Pi}_{2,9}$[6]& 0 & 0&0&$\overline{\Pi}_{3,4}$[12]& 0 & 0&0&$\overline{\Pi}_{4,3}$[6]& 0 & 0&-1/3 \\
$\overline{\Pi}_{2,10}$[12]& 0 &0&1&$\overline{\Pi}_{3,4}$[12]& 0 &0&0&$\overline{\Pi}_{4,4}$[24]& 0 & 0&0 \\
$\overline{\Pi}_{2,11}$[12] & 0 & 0&0&$\overline{\Pi}_{3,5}$[48] & 0 & 0&0&$\overline{\Pi}_{4,4}$[12] & 0 & 1/6&-1/3 \\
$\overline{\Pi}_{2,12}$[24]& 0 & 0&0&$\overline{\Pi}_{3,5}$[24]& 0 & 0&0&$\overline{\Pi}_{4,4}$[24]& 0 & 1/6&0 \\
$\overline{\Pi}_{2,13}$[12]& 0 &1&1&$\overline{\Pi}_{3,5}$[24]& 0 &0&0&$\overline{\Pi}_{4,4}$[48]& 0 & -1/120&0 \\
\end{tabular}
\end{table}

\newpage
   \begin{singlespace}
   \bibliographystyle{model3-num-names_james}
   \bibliography{General_abbrev,Co-superalloys_abbrev}
   \end{singlespace}

\end{document}